\documentstyle[eqsecnum,aps]{revtex}

\begin{document}
\draft
\preprint{hep-th/9805090}
\title{ Adler-Bell-Jackiw anomaly,\\
the Nieh-Yan form and vacuum polarization
}
\author{Chopin Soo\cite{byline1}}
\address{
National Center for Theoretical Sciences\\
P.O. Box 2-131, Hsinchu, Taiwan 300, Taiwan.
}
\maketitle
\begin{abstract}
We show from first principles, using explicitly invariant Pauli-Villars regularization of chiral fermions, that the Nieh-Yan form does contribute to the Adler-Bell-Jackiw (ABJ) anomaly for spacetimes 
with generic torsion, and comment on some of the implications. 
There are a number of interesting and important differences 
with the usual ABJ contribution in the absence of torsion. 
For dimensional reasons, the Nieh-Yan term is proportional 
to the square of the regulator mass. In spacetimes with flat vierbein 
but nontrivial torsion, the associated diagrams are 
actually vacuum polarization rather than triangle diagrams and
the Nieh-Yan contribution to the ABJ anomaly arises from the fact that the axial torsion ``photon" is not transverse.

\end{abstract}

\pacs{Published in Phys. Rev. {\bf D59} (1999) 045006; 
}

\widetext

\section{Preliminaries}

The Adler-Bell-Jackiw (ABJ) anomaly \cite{ABJ} paved 
the way for the elucidation of anomalies in quantum field theories,
and continues to be a fertile link to many diverse topics in
elementary particle physics and gravitation.

Recently, there have been some discussions, and also controversy,
on the question of further contributions to the ABJ anomaly in the presence of 
spacetimes with torsion \cite{Chandia,Obukhov,Kriemer,Guo,Bellisai}. 
We shall show from fundamental principles using Pauli-Villars regularization
that there are indeed further contributions to the ABJ anomaly.
These come in two forms. In addition to the regulator scale 
independent $Tr(\gamma^5 a_2)$ contributions,
there is also the interesting Nieh-Yan term \cite{Nieh-Yan form}
 which diverges as the square of the regulator mass. 
With flat vierbein but non-vanishing
axial torsion, this further ABJ anomaly term is associated with 
vacuum polarization diagrams with two external axial torsion vertices,
rather than with the usual triangle diagrams.

Let us begin by first recalling some basic relations to establish 
the notations. The basic independent ingredients of Riemann-Cartan
spacetimes are the spin connection $A_{AB}$ and the vierbein $e_A$
one-forms.
Lorentz indices are denoted by 
uppercase Latin indices while Greek indices are spacetime indices.
From the definition of the torsion 
\begin{equation}
T_A = de_A + A_{AB}\wedge e^B,
\end{equation}
a generic spin connection can be written (provided the vierbien is invertible)
as the sum of the torsionless spin connection $\omega_{AB}$ 
and terms involving the torsion and vierbein.
Specifically,
\begin{equation}
 A_{\mu AB} = \omega_{\mu AB} - \frac{1}{2}[T_{\sigma\mu A}E^\sigma_B -
T_{\sigma\mu B}E^\sigma_A - T_{\rho\sigma C}e_\mu^C E^\rho_A E^\sigma_B ]
\end{equation}
with $E^\mu_A$ being the inverse of the vierbein $e_{\mu A}$ while
$\omega_{AB}$ satisfies $ de_A + \omega_{AB}\wedge  e^B =0$, and can be
solved as
\begin{eqnarray}
\omega_{\mu AB}& = &\frac{1}{2}
[E^\nu_A (\partial_\mu e_{\nu B} -\partial_\nu e_{\mu B})
- E^\nu_B (\partial_\mu e_{\nu A} -\partial_\nu e_{\mu A})\cr
\nonumber\\
&&-E^\alpha_A E^\beta_B (\partial_\alpha e_{\beta C} 
-\partial_\beta e_{\alpha C})e^C_\mu ].
\end{eqnarray}

Spin $1/2$ fermions couple to torsion through the spin connection 
$\frac{i}{2}A_{\mu AB}\sigma^{AB},$ 
$(\sigma^{AB} = \frac{1}{4}[\gamma^A, \gamma^B])$,
in the covariant derivative
$i{D\kern-0.15em\raise0.17ex\llap{/}\kern0.15em\relax}
= \gamma^\mu(i\partial_\mu + \frac{i}{2}A_{\mu AB}\sigma^{AB}
+ W_{\mu a}{T}^{a})$. Here
$W_{\mu a}$ denotes the generic internal gauge connection in the
${T}^a$ representation.
A number of interesting identities are worth mentioning.
Note that
\begin{eqnarray}
e\frac{i}{2}{\overline{\Psi}}
{A\kern-0.15em\raise0.17ex\llap{/}\kern0.15em\relax}_{AB}\sigma^{AB}\Psi
&=&e\frac{i}{4}A_{\mu AB}(\overline{\Psi}\{\gamma^\mu,\sigma^{AB}\}\Psi +
\overline{\Psi}[\gamma^\mu, \sigma^{AB}]\Psi) \cr
\nonumber\\
&=& \frac{1}{2}(-i A_{\mu AB} J^\nu -\frac{1}{2}\epsilon_{AB}\,^{CD}
A_{\mu CD} J^{5\nu}) E^{\mu B} e^A_\nu
\end{eqnarray}
where $J^\mu = \overline{\Psi}e\gamma^\mu\Psi$ and
$J^{5\mu} = \overline{\Psi}e\gamma^\mu\gamma^5\Psi$ are the densitized 
vector and axial-vector currents.
The anti-commutator term within brackets is 
anti-Hermitian while the commutator term is 
Hermitian. Moreover in Eq. (1.4) the spin connection
coupling (with three $\gamma$-matrices) has been reduced 
to vector and axial-vector couplings. 
In particular, for chiral fermions,
\begin{equation}
e\frac{i}{2}{\overline{\Psi}}_{L,R}
{A\kern-0.15em\raise0.17ex\llap{/}\kern0.15em\relax}_{AB}\sigma^{AB}\Psi_{L,R}
= \frac{1}{2}(-i A_{\mu AB} \pm \frac{1}{2}\epsilon_{AB}\,^{CD}
A_{\mu CD}) E^{\mu B} e^A_\nu J^\nu_{L,R}
\end{equation}
with $J^\mu_{L,R} ={\overline{\Psi}}_{L,R}e\gamma^\mu \Psi_{L,R}
= \mp J^{5\mu}_{L,R}$. This shows that left(right)-handed
chiral fermions couple to the left(right)-handed 
(or anti-self-dual(self-dual)) projection of the spin connection 
 in $i{D\kern-0.15em\raise0.17ex\llap{/}\kern0.15em\relax}$.
By substituting for $A_{\mu AB}$ we may further isolate the 
torsion couplings as
\begin{eqnarray}
e\frac{i}{2}{\overline{\Psi}}_{L,R} 
{A\kern-0.15em\raise0.17ex\llap{/}\kern0.15em\relax}_{AB}\sigma^{AB}\Psi_{L,R}
&=& \frac{1}{2}(-i \omega_{\mu AB} \pm \frac{1}{2}\epsilon_{AB}\,^{CD}
\omega_{\mu CD}) E^{\mu B} e^A_\nu J^\nu_{L,R}\cr
\nonumber\\
&& -(iB_\mu \pm \frac{1}{4e}\tilde{A_\mu})J^\mu_{L,R}
\end{eqnarray}
where $B_\mu \equiv \frac{1}{2} T_{A\mu\nu}E^{\nu A}$ and
$\tilde{A}^\mu \equiv \frac{1}{2}{\tilde\epsilon}^{\alpha\beta\mu\nu}e_{\nu A}
T^A_{\alpha\beta} $ are the trace and axial parts of the torsion
respectively. Similarly, the $B_\mu$ piece is anti-Hermitian while the
term associated with ${\tilde A}_\mu$ is Hermitian.
There are again a few noteworthy remarks.
Both $B_\mu$ and ${\tilde A}_\mu$ are explicitly {\it invariant} under 
local Lorentz transformations while the vierbein and torsion
transform covariantly as rank one Lorentz tensors.
Note that ${\tilde A}_\mu dx^\mu$ is parity-odd (this property
is required for the consistency of the ABJ anomaly equation if the
Nieh-Yan four-form, $d(e_A\wedge T^A)$, contributes to the anomaly).
Indeed ${\tilde A}_\mu dx^\mu$ is the Hodge dual, $*$, of the 3-form 
$e_A \wedge T^A$. Thus its divergence is related to the Nieh-Yan form 
through 
\begin{equation}
\partial_\mu {\tilde A}^\mu =  *d(e_A \wedge T^A).
\end{equation}
All the currents in this article are densitized tensors of weight one.
Thus, for instance, $\partial_\mu J^\mu_L 
= \partial_\mu (e{\overline\Psi}_L\gamma^\mu\Psi_L)$
 is a total divergence while
\begin{equation}
e \nabla_\mu ({\overline\Psi}_L\gamma^\mu\Psi_L) 
=\partial_\mu (e{\overline\Psi}_L\gamma^\mu\Psi_L) - 2eB_\mu{\overline\Psi}_L
\gamma^\mu\Psi_L
\end{equation}
is {\it not} when the torsion trace $B_\mu$ is non-vanishing.
So for the ABJ anomaly, the correct divergence to consider 
for chiral fermions is $\partial_\mu (e{\overline\Psi}_L\gamma^\mu
\gamma^5\Psi_L) =-\partial_\mu (e{\overline\Psi}_L\gamma^\mu\Psi_L)$.

Does $B_\mu$ interact at all with spin $1/2$ chiral fermions?
This depends on whether we couple chiral fermions to gravity 
through the conventional Majorana or Hermitian Weyl prescription, or    
adopt the view that chirality supersedes Hermiticity
and left-handed chiral fermions interact only with the left-handed part 
of the spin connection (for further details, please see \cite{cpt}).
The latter point of view is required by the
(anti)self-dual description of gravity \cite{ash} which extends the  
Weyl nature of the interaction between matter and the forces to 
the gravitational sector \cite{cpt}. In terms of possible ABJ contributions, 
the latter is more general
since $B_{\mu}$ couplings and effects will also be included. 
Thus we shall assume the latter
point of view here and point out the differences with the 
conventional picture so the reader may also deduce what happens then. 
To wit, the bare chiral (Weyl) fermion action is
\begin{equation}
 S = -\int d^4x e{\overline{\Psi}}_{L}i
{{D\kern-0.15em\raise0.17ex\llap{/}\kern0.15em\relax}}P_L\Psi_{L},
\end{equation}
where 
$P_L = \frac{1}{2}(1 - \gamma^5)$ is the left-handed projection operator. 
We adopt the convention
\begin{equation}
\{ \gamma^{A}, \gamma^{B}\}= 2\eta^{AB},
\end{equation}
with $\eta^{AB} = {\rm diag}(-1,+1,+1,+1)$. 
It is clear from the previous comment after Eq. (1.6) that hermitizing
the Weyl action kills the $B_\mu$ coupling completely, and it is for this
reason that one often sees statements to the effect that spin $1/2$ fermions 
interact only with the axial part of the torsion ${\tilde A}_\mu$.

In general, the fermion multiplet $\Psi_{L}$ is in a complex 
representation. This is true of the standard model where there are no 
gauge and Lorentz invariant bare masses. Consequently, this poses a
challenge for the usual invariant Pauli-Villars regularization, even 
though the chiral fermions may belong to an anomaly-free representation.  
An explicitly gauge and Lorentz invariant regularization actually exists
for the standard model; and it can be achieved through an infinite tower
of Pauli-Villars regulators which are doubled in the internal space
(see Ref. \cite{cps} for further details).
Specifically, the internal space is doubled from $T^a$ to
\begin{equation}
{\cal T}^a =\left(\matrix{(-T^a)^* &0\cr 0&T^a}\right),
\end{equation}
and the original fermion multiplet is projected as
$\Psi_{L} = \frac{1}{2}(1-\sigma^3)\Psi_{L}$,
where 
\begin{equation}
\sigma^3 =\left(\matrix{1_d &0\cr 0& -1_d}\right),
\end{equation}
and $d$ is the number of Weyl fermions in the $\Psi_{L}$ multiplet.
With the regularization, we are ready to compute the ABJ anomaly.

\section{The ABJ or $\gamma^5$ ANOMALY}

The ABJ anomaly arises because regularization of the axial vector 
current violates the symmetry of the bare action 
under $\gamma^5$-rotations. This happens for 
instance when Pauli-Villars regularization which maintains gauge invariance
breaks the symmetry through the presence of regulator masses.
We shall calculate the ABJ anomaly first through Pauli-Villars 
regularization of the full standard model and show how it is
related to heat kernel operator regularization. 

Under a singlet chiral $\gamma^5$ rotation, 
\begin{eqnarray}
{\tilde\Psi}_{L} \rightarrow 
e^{i\alpha\gamma^5}{\tilde\Psi}_{L} &=& e^{-i\alpha}{\tilde\Psi}_{L},\cr
\nonumber\\  
{\tilde{\overline\Psi}}_{L} \rightarrow 
{\tilde{\overline\Psi}}_{L} e^{i\alpha\gamma^5}
&=&{\tilde{\overline\Psi}}_{L}e^{i\alpha}. \label{eq:abj}
\end{eqnarray}
For curved spacetimes, we use densitized variables 
${\tilde\Psi}_{L}=e^{1\over 2}{\Psi}_{L},
 {\tilde{\overline\Psi}}_{L}=e^{1\over 2}{\overline\Psi}_{L}.$
The bare massless action is invariant under such a global axial 
transformation, and the associated ABJ or $\gamma^5$ current 
\begin{equation} 
J^\mu_5 = 
{\tilde{\overline\Psi}}_{L}\gamma^\mu\gamma^5\tilde\Psi_{L}
= -{\tilde{\overline\Psi}}_{L}\gamma^\mu\tilde\Psi_{L}
\end{equation}
is conserved classically, i.e. $\partial_\mu J^\mu_5 = 0.$
However, the bare quantum composite current 
\begin{equation}
\langle J^{\mu}_5 \rangle_{bare} =
-\lim_{x \rightarrow y}Tr\left\{\gamma^\mu(x)P_L\left[
\frac{1}{i{{\cal D}\kern-0.15em\raise0.17ex\llap{/}\kern0.15em\relax}}\frac{1}
{2}\left(1-\sigma^3\right)\right]
\delta(x-y)\right\}
\end{equation}
is divergent. The regularized current is however not conserved, as we shall 
show. In Eq. (2.3), (and henceforth 
in $i{D\kern-0.15em\raise0.17ex\llap{/}\kern0.15em\relax}$), 
because of the doubling in internal space, we write the representation 
of the internal gauge field as $W_{\mu a}{\cal T}^a$ in 
${{\cal D}\kern-0.15em\raise0.17ex\llap{/}\kern0.15em\relax}
= e^{1\over 2}{{D}\kern-0.15em\raise0.17ex\llap{/}\kern0.15em\relax}
e^{-{1\over2}}$ and insert the $\frac{1}{2}(1-\sigma^3)$ projection.

As demonstrated in \cite{cps}, 
the expectation value of the Pauli-Villars regularized ABJ current is
\begin{eqnarray}
\langle J^{\mu}_5(x) \rangle_{reg}&=&-\lim_{x \rightarrow y}Tr\{\gamma^\mu(x)
P_L[\frac{1}{2}(1-\sigma^3)(i{{\cal D}\kern-0.15em\raise0.17ex\llap{/}
\kern0.15em\relax})^{\dagger}
\frac{1}{(i{{\cal D}\kern-0.15em\raise0.17ex\llap{/}\kern0.15em\relax})
(i{{\cal D}\kern-0.15em\raise0.17ex\llap{/}\kern0.15em\relax})^\dagger} \cr 
\nonumber\\
&+&\sum_{r=2,4,...}(i{{\cal D}\kern-0.15em\raise0.17ex\llap{/}
\kern0.15em\relax})^{\dagger}
\frac{1}{r^2\Lambda^2 + 
(i{{\cal D}\kern-0.15em\raise0.17ex\llap{/}\kern0.15em\relax})
(i{{\cal D}\kern-0.15em\raise0.17ex\llap{/}\kern0.15em\relax})^{\dagger}}
-\sum_{s=1,3,...}
(i{{\cal D}\kern-0.15em\raise0.17ex\llap{/}\kern0.15em\relax})^{\dagger}
\frac{1}{s^2\Lambda^2 + 
(i{{\cal D}\kern-0.15em\raise0.17ex\llap{/}\kern0.15em\relax})
(i{{\cal D}\kern-0.15em\raise0.17ex\llap{/}\kern0.15em\relax})^{\dagger}}
]\delta(x-y)\} \cr
\nonumber\\
&=&-\lim_{x \rightarrow y}Tr\left\{\gamma^\mu(x)
\frac{1}{2}(1-\gamma^5)
{\frac{1}{i{{\cal D}\kern-0.15em\raise0.17ex\llap{/}\kern0.15em\relax}}}
\frac{1}{2}\left(f({{\cal D}\kern-0.15em\raise0.17ex\llap{/}\kern0.15em\relax}
{{\cal D}\kern-0.15em\raise0.17ex\llap{/}\kern0.15em\relax}^\dagger/\Lambda^2)
-\sigma^3\right)\delta(x-y)\right\}. 
\end{eqnarray} 
where $f(y) = {{(\pi{\sqrt y})}\over{\sinh(\pi{\sqrt y})}}$ is 
the regulator function. So in effect the regulators serve to replace 
the $\frac{1}{2}(1-\sigma^3)$ projection in the bare current by
$\frac{1}{2}[f({{{{\cal D}\kern-0.15em\raise0.17ex\llap{/}\kern0.15em\relax}
{{\cal D}\kern-0.15em\raise0.17ex\llap{/}\kern0.15em\relax}^\dagger}\over 
{\Lambda^2}}) -\sigma^3]$ in the regularized ABJ current. 
The current is regularized for finite values of the regulator mass scale 
$\Lambda$ if $T^a$ is a perturbative anomaly-free representation.
 
The ABJ anomaly 
can be explicitly computed by taking the divergence 
of the expectation value of the regularized expression of Eq. (2.4)
as
\begin{equation}
\langle \partial_\mu J^\mu_5 \rangle_{reg} =
\partial_\mu \lim_{x \rightarrow y}Tr\left\{-\gamma^\mu
\frac{1}{2}(1-\gamma^5)
{\frac{1}{i{{\cal D}\kern-0.15em\raise0.17ex\llap{/}\kern0.15em\relax}}}
\frac{1}{2}\left(f({{\cal D}\kern-0.15em\raise0.17ex\llap{/}\kern0.15em\relax}
{{\cal D}\kern-0.15em\raise0.17ex\llap{/}\kern0.15em\relax}^\dagger/\Lambda^2)
-\sigma^3\right)\delta(x-y)\right\}. 
\end{equation} 
To evaluate the trace, we make use of
the complete sets of 
eigenvectors, $\{{\tilde X}_n\}$ and $\{{\tilde Y}_n\}$, of the 
positive-semidefinite Hermitian operators in Euclidean signature with
\begin{eqnarray}
{{\cal D}\kern-0.15em\raise0.17ex\llap{/}\kern0.15em\relax}
{{\cal D}\kern-0.15em\raise0.17ex\llap{/}\kern0.15em\relax}^\dagger 
{\tilde X}_n &=& 
\lambda^2_n {\tilde X}_n, \cr
\nonumber\\ 
{{\cal D}\kern-0.15em\raise0.17ex\llap{/}\kern0.15em\relax}^\dagger
{{\cal D}\kern-0.15em\raise0.17ex\llap{/}\kern0.15em\relax} 
{\tilde Y}_n &=& 
\lambda^2_n {\tilde Y}_n.
\end{eqnarray}
For the modes with nonzero eigenvalues, ${\tilde X}_n$ and ${\tilde Y}_n$ are 
paired by\footnote{It is assumed that zero modes
have been subtracted from the expectation value of the current. They 
do not occur in the action in the path integral 
formulation \cite{Fujikawa}.}   
\begin{equation}
{\tilde X}_n = {{\cal D}\kern-0.15em\raise0.17ex\llap{/}\kern0.15em\relax} 
{\tilde Y}_n/
\lambda_n, \qquad {\tilde Y}_n ={{\cal D}\kern-0.15em\raise0.17ex\llap{/}
\kern0.15em\relax}^\dagger{{\tilde X}_n}/ \lambda_n.
\end{equation}
Consequently,
\begin{eqnarray}
\langle \partial_\mu J^\mu_5 \rangle_{reg} &=&
-\partial_\mu\left[ \sum_n {\tilde X}^{\dagger}_n{\gamma^\mu} P_L
(i{{\cal D}\kern-0.15em\raise0.17ex\llap{/}\kern0.15em\relax})^\dagger
{\frac{1}{{{\cal D}\kern-0.15em\raise0.17ex\llap{/}\kern0.15em\relax}
{{\cal D}\kern-0.15em\raise0.17ex\llap{/}\kern0.15em\relax}^\dagger}}
\frac{1}{2}
\left(f({{\cal D}\kern-0.15em\raise0.17ex\llap{/}\kern0.15em\relax}
{{\cal D}\kern-0.15em\raise0.17ex\llap{/}\kern0.15em\relax}^\dagger/\Lambda^2)
-\sigma^3\right){\tilde X}_n\right] \cr
\nonumber\\
&=&
i\partial_\mu \left[\sum_n {\tilde X}^\dagger_n{\gamma^\mu}P_L
\frac{1}{2\lambda_n}\left(f(\lambda^2_n/\Lambda^2)
-\sigma^3\right){\tilde Y}_n \right]\cr
\nonumber\\
&=&i[\sum_n \partial_\mu ({\tilde X}^\dagger_n\gamma^\mu)P_L
\frac{1}{2\lambda_n}(f(\lambda^2_n/\Lambda^2) -\sigma^3){\tilde Y}_n \cr
\nonumber\\
&+&\sum_n {\tilde X}^\dagger_nP_R\frac{1}{2\lambda_n}(f(\lambda^2_n/\Lambda^2)
-\sigma^3)\gamma^\mu\partial_\mu {\tilde Y}_n ] \cr
\nonumber\\
&=&-\frac{i}{2}\sum_n[{\tilde Y}^\dagger_n\frac{1}{2}(1-\gamma^5)
(f({{\cal D}\kern-0.15em\raise0.17ex\llap{/}\kern0.15em\relax}^\dagger
{{\cal D}\kern-0.15em\raise0.17ex\llap{/}\kern0.15em\relax}/\Lambda^2)
-\sigma^3){\tilde Y}_n \cr
\nonumber\\
&-& {\tilde X}^\dagger_n\frac{1}{2}(1+\gamma^5)
(f({{\cal D}\kern-0.15em\raise0.17ex\llap{/}\kern0.15em\relax}
{{\cal D}\kern-0.15em\raise0.17ex\llap{/}\kern0.15em\relax}^\dagger/\Lambda^2)
-\sigma^3){\tilde X}_n].
\nonumber\\ \label{eq:abjanomaly}
\end{eqnarray}  
The traces over $\sigma^3$ as well as the parity-even part drop out, and 
the result for Euclidean signature is
\begin{eqnarray}
\langle \partial_\mu J^\mu_5 \rangle_{reg} 
&=&\lim_{\Lambda \rightarrow \infty}
\frac{i}{4}\sum_n[{\tilde Y}^\dagger_n\gamma^5f
({{\cal D}\kern-0.15em\raise0.17ex\llap{/}\kern0.15em\relax}^\dagger
{{\cal D}\kern-0.15em\raise0.17ex\llap{/}\kern0.15em\relax}/\Lambda^2)
{\tilde Y}_n
+ {\tilde X}^\dagger_n\gamma^5f
({{\cal D}\kern-0.15em\raise0.17ex\llap{/}\kern0.15em\relax}
{{\cal D}\kern-0.15em\raise0.17ex\llap{/}\kern0.15em\relax}^\dagger/
\Lambda^2){\tilde X}_n] \cr
\nonumber\\
&=&\lim_{\Lambda \rightarrow \infty}
\frac{i}{4}\sum_n[e{Y}^\dagger_n\gamma^5f
({{D}\kern-0.15em\raise0.17ex\llap{/}\kern0.15em\relax}^\dagger
{{D}\kern-0.15em\raise0.17ex\llap{/}\kern0.15em\relax}/\Lambda^2)
{Y}_n
+ e{X}^\dagger_n\gamma^5f
({{D}\kern-0.15em\raise0.17ex\llap{/}\kern0.15em\relax}
{{D}\kern-0.15em\raise0.17ex\llap{/}\kern0.15em\relax}^\dagger/
\Lambda^2){X}_n] \cr
\nonumber\\
&=& \lim_{\Lambda \rightarrow \infty}\lim_{x \rightarrow x'} 
\frac{i}{4}[ Tr \left\{ e\gamma^5f
({{D}\kern-0.15em\raise0.17ex\llap{/}\kern0.15em\relax}^\dagger
{{D}\kern-0.15em\raise0.17ex\llap{/}\kern0.15em\relax}/\Lambda^2)
{{\delta(x - x')}\over e}I(x,x')\right\}\cr
\nonumber\\
&+& Tr\left\{e\gamma^5f
({{D}\kern-0.15em\raise0.17ex\llap{/}\kern0.15em\relax}
{{D}\kern-0.15em\raise0.17ex\llap{/}\kern0.15em\relax}^\dagger/
\Lambda^2){{\delta(x - x')}\over e}I(x,x')\right\}] .
\end{eqnarray}
We have used 
\begin{equation}
\sum_n e X_n(x) X^\dagger_n(x') = \delta(x - x')I(x,x') ; \qquad 
\sum_n e Y_n(x) Y^\dagger_n(x') = \delta(x - x')I(x,x')
\end{equation} 
with ${\tilde X}_n = e^{1\over 2}X_n $ and ${\tilde Y}_n = e^{1\over 2}Y_n $
and $I(x,x')$ is the displacement bispinor. 

A few remarks are in order.
The operator $i{{D}\kern-0.15em\raise0.17ex\llap{/}\kern0.15em\relax}$ 
is actually not self-adjoint 
in the presence of generic torsion terms. With respect to the 
Euclidean inner product
$\langle W|Z  \rangle = \int_M e W^\dagger Z$, 
\begin{equation}
(i{{D}\kern-0.15em\raise0.17ex\llap{/}\kern0.15em\relax})^\dagger
= e^{-1}iD_\mu \gamma^\mu e = 
i{D\kern-0.15em\raise0.17ex\llap{/}\kern0.15em\relax} + 2iB_\mu \gamma^\mu
\end{equation}
Thus $(i{{D}\kern-0.15em\raise0.17ex\llap{/}\kern0.15em\relax})^2$ 
is not positive-definite, so it is questionable whether regularization
of the axial anomaly with 
$\exp[-(i{D\kern-0.15em\raise0.17ex\llap{/}\kern0.15em\relax})^2/\Lambda^2]$ 
insertion as in Ref.\cite{Chandia} is completely justified in the presence of 
generic torsion which includes nonvanishing $B_\mu$. It is alright
assuming zero $B_\mu$. 
On the other hand, it is clear that regularization 
by the $f({{D}\kern-0.15em\raise0.17ex\llap{/}\kern0.15em\relax} 
{{D}\kern-0.15em\raise0.17ex\llap{/}\kern0.15em\relax}^\dagger 
/\Lambda^2)$ and $f({{D}\kern-0.15em\raise0.17ex\llap{/}
\kern0.15em\relax}^\dagger 
{{D}\kern-0.15em\raise0.17ex\llap{/}\kern0.15em\relax}/\Lambda^2)$ 
pair presented here does not suffer from this defect.
From Eq. (2.11), the {\it self-adjoint} Dirac operator is
$i{{\Delta}\kern-0.15em\raise0.17ex\llap{/}\kern0.15em\relax} 
=i{{D}\kern-0.15em\raise0.17ex\llap{/}\kern0.15em\relax} 
+i{{B}\kern-0.15em\raise0.17ex\llap{/}\kern0.15em\relax} $.
Using this and Eq. (2.11), we have 
\begin{equation}
{{D}\kern-0.15em\raise0.17ex\llap{/}\kern0.15em\relax}^\dagger
{{D}\kern-0.15em\raise0.17ex\llap{/}\kern0.15em\relax}
+{{D}\kern-0.15em\raise0.17ex\llap{/}\kern0.15em\relax}
{{D}\kern-0.15em\raise0.17ex\llap{/}\kern0.15em\relax}^\dagger
= -2 {{\Delta}\kern-0.15em\raise0.17ex\llap{/}\kern0.15em\relax}^2
+ 2 B^\mu B_\mu .
\end{equation}
This relates the
operators on the L.H.S. with the square of the 
self-adjoint Dirac operator. 
Moreover, for Euclidean signature, every term 
in the equation is a positive-definite operator. 
If the Hermitized version of Eq. (1.9) is assumed, the relevant
operator to consider is the square of the self-adjoint
Dirac operator rather than the 
${D\kern-0.15em\raise0.17ex\llap{/}\kern0.15em\relax}^\dagger
{D\kern-0.15em\raise0.17ex\llap{/}\kern0.15em\relax}$ and
${D\kern-0.15em\raise0.17ex\llap{/}\kern0.15em\relax}
{D\kern-0.15em\raise0.17ex\llap{/}\kern0.15em\relax}^\dagger$ pair.

There is an intimate relation between the Pauli-Villars regularization 
presented here and the heat kernel method. This can be seen as follows.
In the form of a power series with Bernoulli numbers $B_k$,
\begin{equation}
\ln({{\sinh y}\over y})= \sum^{\infty}_{k=1} b_k y^{2k}
\end{equation}
with $b_k = {{2^{2k-1}B_{2k}} \over {k(2k)!}}$.
Thus the regulator function takes the form
\begin{equation}
f(y) = \exp(- \sum^{\infty}_{k=1}b_k \pi^k y^k)
\end{equation}
with $y = {{D}\kern-0.15em\raise0.17ex\llap{/}\kern0.15em\relax}
{{D}\kern-0.15em\raise0.17ex\llap{/}\kern0.15em\relax}^\dagger /
\Lambda^2$.
Therefore for $\Lambda \rightarrow \infty$, we may omit terms 
$k >1$ in the series, and the regularization in effect gives the
same result as regularization by $ f({{D}\kern-0.15em\raise0.17ex
\llap{/}\kern0.15em\relax}{{D}\kern-0.15em\raise0.17ex\llap{/}
\kern0.15em\relax}^\dagger/\Lambda^2 ) = 
\exp(-t{{D}\kern-0.15em\raise0.17ex\llap{/}\kern0.15em\relax}
{{D}\kern-0.15em\raise0.17ex\llap{/}\kern0.15em\relax}^\dagger); 
\lim t \rightarrow 0$, with $t =  b_1 \pi/\Lambda^2$.
This allows a direct comparison with heat kernel methods since the operator
$\exp(-t{\hat O})$
satisfies the heat equation
\begin{equation}
-{{\partial K(x,x';t)}\over{\partial t}} = {\hat O}K(x, x';t)
\end{equation}
with $K(x, x';t)= \langle x|\exp(-t{\hat O})|x'\rangle$.

In order to evaluate the ABJ anomaly, 
we have only to compute terms such as 
$\lim_{t \rightarrow 0} \lim_{x \rightarrow x'}
Tr [e\gamma^5 \exp(-t{\hat O})e^{-1}\delta(x -x')I(x, x')]$ 
in Eq. (2.9) for which
 the operator $\hat O$ assumes the form
\begin{eqnarray}
{\hat O} &=& -g^{\mu\nu}D_\mu D_\nu - 2Q^\mu D_\mu + Z\cr
\nonumber\\
&=&- g^{\mu\nu}D'_\mu D'_\nu + X
\end{eqnarray}
where $D'_\mu \equiv D_\mu + Q_\mu$, and $X \equiv D_\mu Q^\mu 
+ Q^\mu Q_\mu + Z$.

The evaluation of anomalies using heat kernel techniques for 
operators of the above form has been pursued in a series of
careful papers by Yajima \cite{Yajima}. We summarize the essential steps
and quote the relevant results.\footnote{In the heat kernel regularization
of Ref.\cite{Yajima}, it is assumed that spinors couple only to the
axial part of the torsion.} 
First we expand $ \delta(x-x')=
\int {{d^4k} \over {(2\pi)}^4}\exp[ik_\mu\sigma^{;\mu}(x,x')]$
with the biscalar $\sigma(x,x')$ being the geodetic interval. 
It is the
generalization of the flat spacetime quantity $\frac{1}{2}(x-x')^2$ for
curved spacetimes, and obeys
\begin{eqnarray}
\sigma(x,x')
& =& \frac{1}{2}g_{\mu\nu}(x)\sigma^{;\mu}(x,x')\sigma(x,x')^{;\nu}\cr
\nonumber\\
& =& \frac{1}{2}g_{\mu'\nu'}(x')\sigma^{;\mu'}(x,x')\sigma^{;\nu'}(x,x').
\end{eqnarray}
The heat kernel of Eq. (2.15) may then be expressed as
\begin{equation}
K(x,x';t) =\int {e{d^4k}\over {(2\pi)^4}}e^{-t{\hat O}}e^{ik_\mu\sigma^{;\mu}}
I(x,x').
\end{equation}
Thus 
\begin{eqnarray}
\lim_{t \rightarrow 0} \lim_{x \rightarrow x'}
Tr [e\gamma^5 \exp(-t{\hat O})e^{-1}\delta(x -x')I(x, x')]
&=& \lim_{t \rightarrow 0} \lim_{x \rightarrow x'}
Tr [\gamma^5 K(x,x';t)]\cr
\nonumber\\
&=&\lim_{t \rightarrow 0}{1\over{(4\pi t)^2}}
 Tr [\sum^\infty_{n=0} e\gamma^5 a_n(x)t^n]
\end{eqnarray}
if we employ the DeWitt ansatz 
\begin{equation}
K(x,x';t) = {1\over{(4\pi t)^2}}
[\det(\sigma^{;\mu\nu}(x,x'))]^{1\over 2}\exp({{\sigma(x,x')}\over {2t}})
\sum^\infty_{n=0} a_n(x,x')t^n
\end{equation} 
which gives the $(x\rightarrow x')$ coincidence limit as
$K(x,x;t) = {e\over{(4\pi t)^2}}\sum^\infty_{n=0} a_n(x)t^n$.
By susbstituting the DeWitt ansatz into the heat equation
and matching the coefficients of powers of t, the recursive relation 
for $a_n$ can be obtained;
from which $a_0 = 1$, $a_1 = (\frac{1}{6}R - X)$ and so on \cite{Yajima}.

The form of the ABJ anomaly is obtained by identifying $Z$ and $Q^\mu$ 
for the specific operators. To wit, 
\begin{eqnarray}
{\hat O}&=&{D\kern-0.15em\raise0.17ex\llap{/}\kern0.15em\relax}^\dagger
{D\kern-0.15em\raise0.17ex\llap{/}\kern0.15em\relax} \cr
\nonumber\\
&=& -g^{\mu\nu}D_\mu D_\nu +
[\Gamma^{\mu}_{\alpha\nu}\gamma^\nu\gamma^\alpha- 2B_\nu\gamma^\nu\gamma^\mu]
D_\mu -\sigma^{\mu\nu}[D_\mu, D_\nu],
\end{eqnarray}
yields
\begin{eqnarray}
-2Q^\mu &=&[\Gamma^{\mu}_{\alpha\nu}\gamma^\nu\gamma^\alpha- 
2B_\nu\gamma^\nu\gamma^\mu], \cr
\nonumber\\
Z &=& -\sigma^{\mu\nu}[D_\mu, D_\nu]\cr
\nonumber\\
&=& -{1\over 2}\sigma^{\mu\nu}\sigma^{AB}F_{\mu\nu AB}
-\sigma^{\mu\nu}G_{\mu\nu a}{\cal T}^a ,
\end{eqnarray}
with $G_{\mu\nu a}$ and $F_{\mu\nu AB}$ being respectively the curvatures of
$W_{\mu a}$ and $A_{\mu AB}$. Similarly, 
\begin{eqnarray}
{\hat O}&=&{D\kern-0.15em\raise0.17ex\llap{/}\kern0.15em\relax}
{{D\kern-0.15em\raise0.17ex\llap{/}\kern0.15em\relax}}^\dagger \cr
\nonumber\\
&=& -g^{\mu\nu}D_\mu D_\nu +
[\Gamma^{\mu}_{\alpha\nu}\gamma^\nu\gamma^\alpha- 2B_\nu\gamma^\mu\gamma^\nu]
D_\mu \cr
\nonumber\\ 
&-&\sigma^{\mu\nu}[D_\mu, D_\nu]-2\gamma^\mu\gamma^\nu\nabla_\mu B_\nu,
\end{eqnarray}
leads to the identification for this latter case of
\begin{eqnarray}
-2Q^\mu &=&[\Gamma^{\mu}_{\alpha\nu}\gamma^\nu\gamma^\alpha- 
2B_\nu\gamma^\mu\gamma^\nu], \cr
\nonumber\\
Z&=& -{1\over 2}\sigma^{\mu\nu}\sigma^{AB}F_{\mu\nu AB}
-\sigma^{\mu\nu}G_{\mu\nu a}{\cal T}^a -2\gamma^\mu\gamma^\nu\nabla_\mu B_\nu
\end{eqnarray}
 
Since $Tr(\gamma^5 a_0) =0$, the first contribution to the ABJ anomaly 
comes from the term proportional to $1/t$ or $\Lambda^2$. 
As dictated by Eq. (2.9), we need to compute 
the sum of the traces of $\gamma^5$ with the $a_1\,'s$ of the two operators 
${D\kern-0.15em\raise0.17ex\llap{/}\kern0.15em\relax}
{D\kern-0.15em\raise0.17ex\llap{/}\kern0.15em\relax}^\dagger$ and 
${D\kern-0.15em\raise0.17ex\llap{/}\kern0.15em\relax}^\dagger
{D\kern-0.15em\raise0.17ex\llap{/}\kern0.15em\relax}$. 
To order $1/t$ or $\Lambda^2$, the result is 
\begin{eqnarray}
\langle\partial_\mu J_5^{\mu}\rangle_{reg.} 
&=&-{i\over 4}(2d){1\over {(4\pi)^2t}}
[-2{1\over 2}{\tilde \epsilon}^{\alpha\beta\mu\nu}e^A_\alpha 
e^B_\beta F_{\mu\nu AB}
+ 2(g_{\mu\nu}{\tilde \epsilon}^{\alpha\eta\beta\rho}
\Gamma^\mu_{[\alpha\eta]}\Gamma^\nu_{[\beta\rho]})]\cr
\nonumber\\
&=&-{{i\times d}\over {(4\pi)^2t}}[-{1\over 2}
{\tilde \epsilon}^{\alpha\beta\mu\nu}e^A_\alpha e^B_\beta F_{\mu\nu AB}
+ {1\over 4}{\tilde \epsilon}^{\alpha\beta\mu\nu}T_{\alpha\beta A} 
T^A_{\mu\nu})]\cr
\nonumber\\
&=&-{{i\times d}\over {(4\pi)^2t}} *[-e^A\wedge e^B\wedge F_{AB}
+  T^A \wedge T_A]\cr
\nonumber\\
&=&-{{i\times d}\over {(4\pi)^2t}}*[d(e^A\wedge T_A)].
\end{eqnarray}
In the above, we have used 
$\Gamma^\mu_{[\alpha\beta]}= -{1\over 2}E^{\mu A}T_{\alpha\beta A}$ while
$d(e^A\wedge T_A)$ is precisely the Nieh-Yan four-form \cite{Nieh-Yan form}. 
We have therefore confirmed from first principles using Pauli-Villars 
regularization of chiral fermions that the Nieh-Yan four-form 
indeed contributes to the ABJ anomaly \cite{Chandia}. 
In the absence of torsion, 
the $\Lambda$-independent terms from $Tr(\gamma^5 a_2)$ 
are the familiar curvature-squared ABJ contributions. 
Using the heat kernel method, Obukhov et al \cite{Obukhov} have also found
the Nieh-Yan contribution to the ABJ anomaly using the operator
$ {\hat O} = -{{\Delta}\kern-0.15em\raise0.17ex\llap{/}\kern0.15em\relax}^2$.
It is implicitly present in $Tr(\gamma^5 a_1)$ in the series of papers by 
Yajima \cite{Yajima}, and also in Ref.\cite{Nieh-Yan}.

\section{Further remarks}

The Nieh-Yan term is proportional
$1/t$ and hence to the square of the Pauli-Villars regulator 
mass scale $\Lambda^2$ for dimensional reasons.
Moreover, it is also clear from the discussion here that the Nieh-Yan
contribution is indeed due to regularization and is compatible with
the general understanding of the origin of anomalies in quantum field theories.
It must be stressed that we do not have to make statements to the effect
that the integration over $k$ in Eq. (2.18) is truncated at some 
scale $M_{cut-off}$ so that $\int d^4k \approx M_{cut-off}^4$ and so on. 
These statements have the essence of introducing an extra cut-off scale 
$M_{cut-off}$ which may or may not be $\Lambda$; and can lead to confusion 
on the dependence of divergent ABJ contributions on the regulator scale 
$\Lambda$. The cut-off is unnecessary because the integrals are really 
well-defined due to the presence of the regulating function f and its 
derivatives (for instance, $\int d^4k \exp(-k^2/\Lambda^2) = \pi^2 \Lambda^4$).
The upper limit of the integrals is really $\infty$; and no extra
cut-off mass scale is needed. There is only one regulator mass 
scale $\Lambda$.

In covariant operator regularization, a regulating function
$f({\hat O})$ is inserted in the trace with $\gamma^5$ in Eq. (2.9).
This does not necessarily imply that the anomaly is independent of ${\hat O}$
although it is independent of the specific form of $f$
with the same boundary conditions, as has been emphasized by 
Fujikawa \cite{Fujikawa}. 
There may still be some leeway and ambiguity in selecting 
the operator ${\hat O}$.
In our case, the operator in Eq. (2.9) is dictated
by the requirement that to regularize gauge and spin currents 
and also the energy-momentum of the full theory using 
the generalized Pauli-Villars method \cite{cps}, it is essential that 
the regulators couple to chiral fermions in the same manner as specified 
by the bare Lagrangian.

In dimensional regularization of fermion loops, there is the subtlety with
$\gamma^5$ leading to inconsistencies 
(see for instance  Ref. \cite{Collins} and references therein) if 
$\{\gamma^\mu, \gamma^5\} =0$ is maintained for arbitrary values of the
spacetime dimension. A consistent set of relations is 
$\{\gamma^5, \gamma^\mu \}=0$ if $\mu =0,1,2,3$; $[\gamma^5, \gamma^\mu] =0$
otherwise. This implies that global $\gamma^5$ rotations no longer generate 
symmetries if the dimension of spacetime is not 4. Consequently,
in the axial vector current we have 
$\partial_\mu J^\mu_5 = {\overline{\Psi}}e\gamma^5 
i{D\kern-0.15em\raise0.17ex\llap{/}\kern0.15em\relax}\Psi$.
On taking the expectation value of this equation, we see the ABJ 
anomaly as the expectation value of the R.H.S. 

Mielke and Kriemer \cite{Kriemer} have argued that the Nieh-Yan
form cannot contribute to the ABJ anomaly because perturbatively 
it cannot come from triangle diagrams. The first part of the argument 
is incorrect and it is therefore instructive to see what
 Feynman diagram processes are associated with the Nieh-Yan contribution. 
Things are much clearer if we specialize to flat vierbein 
$e^A\,_\mu = \delta^A\,_\mu$, but with non-trivial axial torsion 
 ${\tilde A}_\mu$. This allows us to retain the essential
information regarding the Nieh-Yan contribution without having to
worry about background graviton fluctuations from $e_{A\mu}$.
Let us also set $W_{\mu a} = 0$ and $B_\mu =0$ for convenience. 
Then the action of Eq. (1.9) reduces to
\begin{equation}
S= -\int d^4x {\overline{\Psi}_L}\gamma^\mu(ie\partial_\mu 
- \frac{1}{4}{\tilde A}_\mu)\Psi_L ,
\end{equation}
and the torsion coupling is ``QED-like". It is also clear 
that the ABJ current $J^\mu_5$ is coupled to $\tilde A_\mu$. 
Thus for chiral fermions the ABJ current is the source for axial torsion. 
In Pauli-Villars regularization, fermion loops with
background ${\tilde A}_\mu$ vertices
are obtained by functionally differentiating the regularized current with
respect to ${\tilde A}_\mu$.   
As computed, the Nieh-Yan contribution (after continuation to Lorentzian 
signature) is
\begin{equation}
\partial_\mu \langle J^\mu_L \rangle 
= \frac{d}{(4\pi)^2t}\partial_\mu {\tilde A}^\mu.  
\end{equation}
This implies
\begin{equation}
\partial_\mu 
{{\delta\langle J^\mu_L(x)\rangle}\over {\delta {\tilde A}_\nu (x')}}
=\frac{d}{(4\pi)^2t}\partial^\nu \delta(x-x').
\end{equation}
But the vacuum polarization amplitude $\Pi^{\mu\nu}$ with 
two external background ${\tilde A}$ 
vertices is proportional to the Fourier transform of 
the functional derivative of the current with respect to ${\tilde A}$,
i.e.
\begin{equation}
\left.{{\delta\langle J^\mu_L(x)\rangle}\over {\delta {\tilde A}_\nu (x')}}
\right|_{{\tilde A}=0}
\propto \int {{d^4k} \over {(2\pi)^4}}e^{ik.(x-x')}\Pi^{\mu\nu}.
\end{equation}
In momentum space, Eq. (3.3) means that the Ward identity which corresponds to 
the Nieh-Yan contribution of the ABJ anomaly reads 
\begin{equation}
k_\mu \Pi^{\mu\nu} \propto k^\nu .
\end{equation}
This is consistent with 
\begin{equation}
\Pi^{\mu\nu} = (g^{\mu\nu} - (k^\mu k^\nu/k^2)) \Pi(k^2) + \Pi' g^{\mu\nu}
\end{equation} 
If $\Pi'=0$, we recover then the usual ``transverse photon" condition of
``gauge" invariance i.e.
$k_\mu \Pi^{\mu\nu}= 0$ and $\langle \partial_\mu J^\mu_L\rangle =0$
for the vacuum polarization diagram.
However, we must remember that ${\tilde A}_\mu$ is {\it not} a gauge
potential, but a composite; and is completely invariant 
(as emphasized in Section I) under local Lorentz transformations 
which are actually gauged with the spin connection $A_{\mu AB}$.
Thus local Lorentz invariance is {\it not} anomalous
as evidenced by the explicitly Lorentz (and also gauge) invariant 
regularization scheme \cite{cps}. 
Even if we include the full $W_{\mu a}T^a$ and $B_\mu$
couplings, there are no perturbative chiral {\it gauge} 
anomalies provided $Tr (T^a)=Tr(T^a\{T^b, T^c\})= 0$ \cite{cps,Glash,nieh}.
However, the current that is coupled to the parity-odd ${\tilde A}$ 
composite is none other than the ABJ current
which is anomalous (see Eq. (3.2))
because the ``photon" ${\tilde A}_\mu$ is not transverse
i.e. $\partial_\mu {\tilde A}^\mu \neq 0$ precisely when the Nieh-Yan form
is non-vanishing.

Since ${\tilde A}^\mu$ is actually local Lorentz invariant and transforms
covariantly as a general coordinate tensor density, it may be
possible to redefine the current (and the corresponding charge) 
generating axial rotations by $J^{\mu}_5 + {1\over{(4\pi)^2t}}{\tilde A}^\mu$.
This is in contradistinction with the 
usual ${\tilde \epsilon}^{\alpha\beta\mu\nu} G_{\alpha\beta a} G_{\mu\nu}^a$ 
contribution for which we cannot construct a gauge invariant physical 
current by absorbing the gauge-dependent Chern-Simons current 
(similarly, in the gravitational case, the associated Chern-Simons current 
also does not transform covariantly under general coordinate transformations). 
However, absorbing ${\tilde A}^\mu$ into $J^{5\mu}$ can lead to 
interesting changes in the scaling behaviour of the redefined current and 
the renormalization properties. These are currently under investigation. 
It is also clear that $e_A\wedge T^A$
is local Lorentz invariant and globally defined (even if the vierbein 
and spin connections are defined only locally), in the sense that
in the overlap of patches 1 and 2, $(e_A)_1 \wedge T^A_1 
= (e_A)_2 \wedge T^A_2$. So the Nieh-Yan term of the ABJ anomaly 
 gives zero contribution when integrated over compact manifolds 
without boundaries \cite{Guo}, and therefore does not affect the 
Atiyah-Singer index theorem in such cases. However for manifolds 
with boundaries, $\int_{\partial M} e_A \wedge T^A$ can be 
non-trivial \cite{Chandia,Obukhov}. 
So axial rotations of fermions in the presence of torsion will then lead to
extra P, CP and T violations from Nieh-Yan ABJ contributions
over and beyond the usual instanton terms.

\acknowledgments
The research for this work has been supported by funds from
the National Center for Theoretical Sciences, Taiwan, under
Grant No. NSC87-2119-M-007-004.
C.S. would like to thank L. N. Chang for helpful correspondence
and Darwin Chang for a useful conversation.


\end{document}